\documentclass[structabstract]{aa}
\usepackage{natbib}
\bibpunct{(}{)}{;}{a}{}{,}
\usepackage{graphicx}
\usepackage{txfonts}
\usepackage{rotating}

\begin{document}
\title{A Comprehensive Study of Large Scale Structures in the GOODS-SOUTH Field up to $z \sim 2.5$}
   \author{S. Salimbeni \inst{1,2}
   \and
    M. Castellano \inst{3,2}
\and
    L. Pentericci
    \inst{2}
\and
    D. Trevese
    \inst{3}
\and
    F. Fiore
     \inst{2}
  \and
  A. Grazian
    \inst{2}
 \and
    A. Fontana
     \inst{2}
   \and
   E. Giallongo
    \inst{2}
    \and
     K. Boutsia 
    \inst{2}
   \and
   S.Cristiani
    \inst{2}
   \and
   C. De Santis
    \inst{5,6}
   \and
   S. Gallozzi
    \inst{2}
   \and
   N. Menci
    \inst{2}
\and
    M. Nonino
    \inst{4}
 \and
D. Paris
 \inst{2}
   \and
    P. Santini
     \inst{2}
   \and 
    E. Vanzella
    \inst{4}
   }

   \offprints{S. Salimbeni, \email{salimben@astro.umass.edu}}

\institute{Department of Astronomy, University of Massachusetts, 710 North Pleasant Street, Amherst, MA 01003 \and  INAF - Osservatorio Astronomico di Roma, Via Frascati 33,
I--00040 Monteporzio (RM), Italy \and Dipartimento di Fisica,
Universit\'{a} di Roma ``La Sapienza'', P.le A. Moro 2, 00185 Roma,
Italy \and INAF - Osservatorio Astronomico di Trieste, Via G.B. Tiepolo 11, 34131 Trieste, Italy \and Dip. di Fisica, Universit\'{a}  Tor Vergata,
Via della Ricerca Scientifica 1,
00133 Roma, Italy  \and INFN-Roma Tor Vergata,
Via della Ricerca Scientifica 1,
00133 Roma, Italy}

   \date{Received .... ; accepted ....}
   \titlerunning{Large Scale Structures in the GOODS-SOUTH Field}
  
\abstract
{}
{
The aim of the present paper is to identify and study the properties and galactic content of groups and  clusters in the GOODS-South field up to $z\sim
 2.5$, and to analyse the  physical properties of galaxies as a continuous function of environmental density up to high redshift.}
{We use the deep ($z_{850}\sim 26$), multi--wavelength GOODS-MUSIC
catalogue, which has a 15\% of spectroscopic redshifts and accurate
photometric redshifts for the remaining fraction. On these data, we
apply a (2+1)D algorithm, previously developed by our group, that provides an adaptive estimate of the 3D density field.
 We support our analysis with simulations to
evaluate the purity and the completeness of the cluster catalogue
produced by our algorithm.}
{We find several high density peaks embedded in larger structures
in the redshift range 0.4-2.5. From the analysis of their physical
properties (mass profile, $M_{200}$, $\sigma_v$, $L_X$, $U-B$ vs.
$B$ diagram), we derive that most of them are groups of galaxies,
while two are poor clusters with masses of few times $10^{14}
M_\odot$. For these  two clusters we find, from the Chandra 2Ms data, an X-ray emission significantly lower than expected from
their optical properties, suggesting that the two clusters are
either not virialised or gas poor. We find that the slope of the
colour magnitude relation, for these groups and clusters, is constant at least up
to $z \sim 1$. We also analyse the
dependance on environment of galaxy colours, luminosities, stellar
masses, ages and star formations. We find that galaxies in high density regions are,
on average, more luminous and massive than field galaxies up to $z
\sim 2$. The fraction of red galaxies increases with luminosity
and with density up to $z\sim 1.2$. At higher $z$ this dependance
on density disappears. The variation of galaxy properties as a
function of redshift and density suggests that a significant change
occurs at $z \sim 1.5-2$.}
{}


\maketitle
%
\section{Introduction}

The study of galaxy clusters and of the variation of galaxy properties as a function of the environment
are fundamental tools to understand the formation and evolution  of the large scale structures and of the different galaxy populations, observed both in the local
and in the high redshift Universe.  The effects  of the environment on galaxy evolution have been studied at progressively higher redshifts through the analysis
of single clusters \citep[e.g.][]{Treu03,nakata2005,tran2005,Mei06,Menci08,Rettura08}, as well as studying the variation of galaxy colours, morphologies and other
physical parameters as a function of projected or 3-dimensional density  \citep[e.g.][]{dress2,Blanton05,cucciati06,cooper2007,Elbaz07}. Moreover, the analysis of cluster properties at different wavelenghts   provides interesting insights into the matter content and evolutionary histories of these structures  \citep{Lubin2004,rasmussen2006,popesso2007}.

A variety of survey techniques have proved effective at finding
galaxy clusters up to z$\sim$ 1 and beyond. X-ray selected samples
at z $\ge$ 1  probe the most massive and dynamically relaxed systems
\citep[e.g., ][]{Maughan2004,stanford2006,bremer2006, lindman2008}. Large-area
multicolour surveys, such as the red-sequence survey
\cite[e.g.][]{gladders2005}, have collected  samples of systems in a
range of evolutionary stages. The mid-IR cameras on board of the Spitzer and Akari
satellites  has extended the range and power of multicolour surveys,
producing confirmed and candidate clusters up to  $z \sim 1.7$
\citep{stanford2005, eisenhardt08,goto2008}. However most of the previous techniques
present some difficulties in the range $1.5<z<2.5$, where we expect
to observe the formation of the red sequence and the first hints of
colour segregation \citep{cucciati06,kodama2007}. Searching  for extended X-ray
sources becomes progressively more difficult at large distances,
because the surface brightness of the X-ray emission fades as
$(1+z)^4$.  The sensitivity of surveys exploiting the Sunyaev-Zeldovich (SZ) effect is, at present,  not sufficient to detect any of known clusters at $z>1$  
\citep{carlstrom2002,staniszewski2008}. Finally, the detection of galaxy overdensities on surveys using  two-dimensional algorithms, requires additional {\it a priori} assumptions on  either
galaxy luminosity function (LF), as in the Matched Filter algorithm \citep{postman1996}, or relies on the presence of a red sequence \citep{gladders2000}. Biases produced by these assumptions can
hardly be evaluated at high redshift. 

In this context, photometric redshifts obtained from deep multi-band
surveys for large samples of galaxies, though having a relatively
low accuracy  if compared to spectroscopic redshifts, can be exploited to detect
and study distant structures. In the past few years, several authors
\citep[e.g.][]{botzler04,vanbreukelen07,scoville07,zatloukal07,mazure07,eisenhardt08}
have developed or extended known algorithms to take into account the
greater redshift uncertainties. We have developed a new algorithm,
that uses an adaptive estimate of the 3D density field, as described
in detail in \cite{trevese2006}. This method combines galaxy angular
positions and precise photometric redshifts  to estimate the galaxy
number-density and to detect galaxy over-densities in three
dimensions also at $z>1$, as described in Sect. 3.

Our first application to the K20 survey \citep{cimatti2002} detected
two clusters at z$\sim$0.7 and z$\sim$1 \citep{trevese2006},
previously identified through spectroscopy \citep{gilli2003,adami2005}.
We then applied the algorithm to the much larger GOODS-South field,
and in \cite{castelllano2007}, elsewhere C07, we reported our
initial results, i.e. the discovery of a forming cluster of galaxies
at z$\sim 1.6$.

In this paper we present the application of the algorithm to the
entire GOODS-South area ($\sim 143 \ arcmin^2$), using the
GOODS-MUSIC catalogue \citep{grazian} up to $z \sim 2.5$, to give a
comprehensive description of the large scale structures in this
field, with a detailed analysis of the  physical properties of each
high density peak. We also study the physical properties of galaxies
as a function of environmental density up to redshift 2.5, 
higher than previous similar studies \citep[e.g.,][]{cucciati06,cooper2007}.

To validate our technique, we analysed the completeness and
purity of our cluster detection algorithm, up to $z\sim  2.5$,
through its application to a set of numerically simulated galaxy
catalogues. Besides allowing an assessment of the physical reality
of the structures found in the GOODS field, this analysis provides
the starting point to test the reliability of the algorithm in view
of our plan to apply it to photometric surveys of similar depth but
covering much larger areas.

The paper is organised as follows: in Sect. 2, we describe the basic
features of our dataset. In Sect. 3, we summarise the basic features
of the (2+1)D algorithm used in our analysis, and compare it with
other methods based on photometric redshifts. In Sect. 4, we show
the results of the application of our method to simulated data. In Sect. 5, we present the
catalogue of the structures detected and the derived physical
properties. In Sect. 6, we study the colour magnitude diagrams of
the detected structures. In Sect. 7, we analyse the physical
properties of galaxies as a continuous function of environmental
density.

All the magnitudes used in the present paper are in the AB system,
if not otherwise declared. We adopt a cosmology with  $\Omega_\Lambda=0.7$, $\Omega_M=0.3$,
and $H_0=70$ km s$^{-1}$ Mpc$^{-1}$.

\section{The GOODS-MUSIC catalogue}\label{catalog}

We used the multicolour  GOODS-MUSIC catalogue \citep[GOODS MUlticolour Southern Infrared Catalogue;][]{grazian}.  This catalogue comprises information in 14 bands (from U band to 8$\mu m$) over an area of about 143.2 $arcmin^2$.  We used the $z_{850}$-selected sample ($z_{850} \sim 26$), that  contains
9862 galaxies (after excluding AGNs and galactic stars). About 15\% of the galaxies
in the sample have spectroscopic redshift, and for the other galaxies we
used  photometric redshifts  obtained  from a standard $\chi^2$ minimisation over a large set
of spectral models \citep[see e.g.,][]{fontana00}.  The accuracy of the
photometric redshift is very good, with a r.m.s. of 0.03 for the $\Delta z/(1+z)$ distribution up to redshift $z=2$. For a detailed description of the catalogue we refer to \citet{grazian}.

The method we applied  to estimate the rest-frame magnitudes and the
other physical parameters (M, SFR, age) is described in previous papers
\citep[e.g., ][]{fontana06}. Briefly, we use a $\chi^2$ minimisation
analysis, comparing the observed SED of each galaxy to synthetic
templates, and the redshift is fixed during the fitting process to
the spectroscopic or photometric redshift derived in
\citet{grazian}. The set of templates is computed with standard
spectral synthesis models \citep{bruzual2003}, chosen to broadly
encompass the variety of star formation histories, metallicities
and extinctions of real galaxies. For each model of this grid, we
 compute the expected magnitudes in our filter set, and find
the best--fitting template. From the best--fitting template we
obtain, for each galaxy, the physical parameters that we use
in the analysis. Clearly, the physical properties are subject
to uncertainties and biases related to the synthetic libraries used
to fit the galaxy SEDs; however, as shown in \cite{fontana06}, the
extension of the SEDs to mid-IR wavelengths with IRAC tends to
reduce the uncertainties on the derived stellar masses. 
For a detailed analysis of the uncertainties on the physical properties we
refer to our previous papers \cite[][]{fontana06,grazianeros}.
In the present work we also make use of the 2Ms X-ray observation of the Chandra Deep Field South 
presented by \citet{luo2008} and of the catalogue of VLA radio sources (1.4 GHz) on the CDFS compiled by \citet{miller2008}.

\section{The (2+1)D algorithm for the density estimation}\label{algorithm}

To estimate a three dimensional density, we developed a method that
combines the angular position with the photometric redshift of each
object. The algorithm is described in detail in \citet{trevese2006}:
here we outline its main features and, in the next section, we
present the simulations used to estimate its reliability.

The procedure is designed to automatically take into account the
probability that a galaxy in our survey is physically  associated to a given overdensity.  This
is obtained by computing  the galaxy densities in volumes whose shape is
proportional to positional uncertainties in each dimension ($\alpha$, $\delta$  and $z$).

First, we divide   the volume of the survey in cells whose extension in
different  directions ($\Delta\alpha, \Delta \delta, \Delta z$)
depends on the relevant positional accuracy and thus are elongated
in the radial direction. We choose the cell sizes small enough to
keep an acceptable spatial resolution, while avoiding a useless
increase of the computing time. We  adopt $\Delta z =0.025$
(radial direction) and $\Delta \alpha = \Delta\delta  \sim2.4\
arcsec$ in transverse direction, the latter value corresponds to $\sim 30$, 40 and $60 kpc$ (comoving),  respectively at $z\sim 0.7$, $\sim 1.0$ and
$\sim 2.0$. 

For each cell in space we then count neighbouring
objects in volumes that are progressively increased in each direction by steps of one
 cell, thus keeping the simmetry imposed by the different intrinsic
 resolution. When a number $n$ of objects is
reached we assign to the cell a comoving density $\rho = n/
V_n$, where $V_n$ is the comoving volume which includes the
n-nearest neighbours. 
Clusters would be better characterized by their proper density since they 
already decoupled from the Hubble flow, however we notice that the average uncertainty on photometric redshifts, that grows with redshift as $(1+z)$, forces us to measure densities in volumes that are orders of magnitude larger than the real volume of a cluster, even at low-$z$.  Thus we decided to measure comoving densities, that have the further advantage of giving a redshift-independent density scale for the background.
We fix $n = 15$ as a trade off between
spatial resolution and signal-to-noise ratio. 
Indeed, through simulation described in Sect. \ref{simulazioni}, we verified that a lower  $n$ would greatly raise the high frequency
noise in the density maps, thus increasing the contamination from false detections  
in the cluster sample, even at low redshift ('purity' parameter in Tab. \ref{tab:simulazioni1}).
In the density estimation, we assign a weight $w(z)$ to each
detected galaxy at redshift $z$, to take into account the increase
of limiting absolute magnitude with increasing redshift for a given
apparent magnitude limit. We choose $w(z)=1/s(z)$, where $s(z)$ is
the fraction of objects detected with respect to a reference
redshift $z_c$ below which we detect all objects brighter than the
relevant $M_c \equiv M_{lim}(z_c)$:
\begin{equation}
s(z)=\frac{\int^{M_{lim(z)}}_{-\infty}\Phi(M)dM}{\int^{M_c}_{-\infty}\Phi(M)dM},
\end{equation}
where $\Phi(M)$ is the redshift dependent galaxy luminosity function
computed on the same GOODS-MUSIC catalogue \citep{salimbeni2008},
 $M_{lim}(z)$ is the 
absolute magnitude limit at the given redshift $z$, corresponding to the  apparent magnitude limit $m_{lim}$ of the survey, which depends on the
position \citep[see ][]{grazian}. We use this
correction to obtain a density scale independent of redshift, at
least to a first approximation. In computing $M_{lim}(z)$, we use K
- and evolutionary - corrections for each object computed with the
same best fit SEDs used to derive the stellar masses, the
rest--frame magnitudes, and the other physical properties.

We apply this algorithm  to data from the GOODS-MUSIC catalogue, in
a redshift range from $z\sim0.4$ to $z\sim 2.5$, where we have
sufficient statistic. We perform this analysis
selecting galaxies brighter than $M_B=-18$ up to redshift 1.8 and
brighter than $M_B=-19$ at higher redshift, to minimise the
completeness correction described above, keeping the average
weight $w(z)$ below 1.6 in all cases.

Using this comoving density estimate we analyse the field in two
complementary ways. First, we detect and study galaxy
overdensities, i. e. clusters or groups  (see Sect. \ref{sec:overdens}), defined as
connected 3-dimensional  regions with density exceeding a fixed
threshold and a minimum number of members chosen according to the results of the simulations  (Sect.
\ref{simulazioni}). In particular, we isolate the structures as the regions having
$\rho>\bar{\rho}+4 \sigma$ on our density maps and at least 5 members. We then consider as
part of each structure the spatially connected region (in RA,
DEC, and redshift) around each peak, with an environmental density
of $>2\sigma$ above the average and at least 15 member galaxies. To avoid spurious connections
between different structures at the same redshift, we consider
regions within an Abell radius from the peak. The galaxies located
in this region are associated with each structure.  We then study the variation of galaxy
properties as a function of environmental density  (Sect.
\ref{sec:fieldprop}), associating to each galaxy in the sample the comoving density at its position.

\subsection{Comparison with other methods based on photometric redshifts}

 As mentioned  in the introduction, other methods, based on
photometric redshifts, have been developed for the detection of
cosmic structures.  Here we present the main differences between our
algorithm and those which appeared most recently in the literature.
However, a more detailed comparison is beyond the scope of the
present work, since it would require extensive simulations and/or the
application of the different methods to the same datasets.

A similar  three-dimensional approach has been proposed by
\citet{zatloukal07}. They select cluster candidates  detecting excess density
in the 3D galaxy distribution reconstructed from the photometric redshift
probability distributions $p(z)$ of each object. However, at variance with our method they do not adopt any redshift dependent correction for their
estimated density, since they analyze only a small
redshift range. As we outlined in the previous section, such correction is needed  to
provide a redshift independent density scale in a deep sample as the GOODS-MUSIC.

 \citet{botzler04} expanded the well known Friends of Friends (FoF)
algorithm \citep{Huchra82}, to take into account photometric redshift uncertainties.
This method links together groups of individual galaxies if their
redshift difference and angular distances are below fixed
thresholds. These thresholds depend on the photometric redshift
uncertainties, which are greater than the average physical distance
between galaxies and also greater than the velocity dispersion of
rich clusters. This could induce the problem of structures
percolating through excessively large volumes. They dealt with this
issue dividing the catalogue in redshift slices.  Instead of
comparing the distance between galaxy pairs, as done in a FoF
approach, we use the statistical information of how many galaxies
are in the neighbourhood  of a given point to estimate a physical
density.  This approach can avoid more effectively the percolation
problem, since it identifies structures from the 4$\sigma$  density peaks
whose extension is limited by the fixed threshold in density. 

Several authors, e.g. \citet{scoville07}, \citet{mazure07}, \citet{eisenhardt08}  and \citet{vanbreukelen07}
estimated the  surface density in redshift
slices, each with different methods: the first two use adaptive
smoothing of galaxy counts, \citet{eisenhardt08} analyse a density map convolved with a wavelet kernel,  while the last author adopts FoF and
Voronoi tessellation \citep{Marinoni02}.  At variance with these, we prefer to adopt an
adaptive 3D density estimate to consider, automatically, distances in all directions and the relevant positional accuracies at the same time. This approach
requires longer computational times, but  allows for an
increased resolution in high density regions where the chosen number
of objects is found in a smaller volume with respect to field and
void regions. As a consequence it also avoids all peculiar ``border''
effects given by the limits of the redshift slices, and there is
also no need to adopt additional criteria  to decide whether an
overdensity, present in two contiguous 2D density maps in similar
angular positions, represents the same group or not   \citep[as done
for example by][]{mazure07}. This clearly also depends on the
ability of the algorithm to separate aligned structures (for a more
detailed discussion of this see Sect. \ref{simulazioni}).
 
Finally, another important difference with respect to previous
methods is in the way we use the photometric redshift: some authors
used best fit values of photometric redshift, e.g. \cite{mazure07},
while \cite{scoville07}, \cite{vanbreukelen07}, \citet{zatloukal07} and
\citet{eisenhardt08} consider the full probability
distribution function (PDF) to take into account redshift
uncertainties. As discussed by \citet{scoville07}, this last method
could tend to preferentially detect structures formed by early type
galaxies, since they have smaller photometric redshift uncertainty, thanks
to their stronger Balmer break, when this feature is well sampled in
the observed bands. We are less biased in  this respect, since we consider the photometric
redshift uncertainty in a conservative way,  choosing only the
maximum redshift range  where we count neighbour galaxies to associate
with each cell. We took this range as $\pm 2\cdot \sigma_z$ around
the redshift of each cell, where $\sigma_z=0.03 \cdot (1+z)$
\citep{grazian} is the average accuracy of the  photometric
redshift in the range we analyze.

\section{Simulations}\label{simulazioni}

We estimate the reliability of our cluster detection algorithm by
testing it on a series of mock catalogues, designed to reproduce the
characteristics of the GOODS survey. These mock catalogues are
composed by a given number of groups and clusters superimposed on a
random (poissonian) field. While this is a rather simplistic
representation of a survey, it allows us to evaluate some basic
features of our algorithm, without the use of N-body simulations. We
expand the previous simulations presented in \cite{trevese2006},
using a larger number of mock catalogues and adopting a more
consistent treatment of the survey completeness. 
For each redshift, we calculate the limiting absolute B magnitude for the two populations of
``red'' and ``blue'' galaxies, defined from the minima in the U-V vs. B distribution in \cite{salimbeni2008}, using the average type-dependant K- and
evolutionary corrections calculated from the best-fit SED of the
objects in the real catalogue. We then generate an ``observed''
mock catalogue of field galaxies randomly distributed over an area  equal to that of the GOODS-South survey.  At each redshift, the number of objects in the catalogue is obtained  from the 
integral of the rest frame $B$ band luminosity function
$\Phi(M_B,z)$ derived in \cite{salimbeni2008}, up to the limiting absolute $M_B(z)$ magnitude computed as described above.

\begin{table}[!h]
\caption{Completeness and Purity} \label{tab:simulazioni1}
\begin{center}
\begin{tabular}{lrrrr}
\hline \noalign{\smallskip} \hline\noalign{\smallskip} Mass & Purity
& Un.
Pairs  &  Compl. & Double Id.   \\
\noalign{\smallskip}\hline\noalign{\smallskip}
& \multicolumn{4}{c}{$0.4<z<1.2$}\\
\noalign{\smallskip}\hline\noalign{\smallskip}
M$>1\times 10^{13} \ M_{\odot}         $  &   100\% & 6.5\%  &  86.2\%    &  4.2\%  \\
M$>2 \times 10^{13} \ M_{\odot} $  &   100\% & 3.1\%  &  89.7\%    &  0\%    \\
M$>3 \times  10^{13} \ M_{\odot}$  &  93.8\% &   0\%  &  93.8\%    &  0\%    \\
\noalign{\smallskip}\hline\noalign{\smallskip}
& \multicolumn{4}{c}{$1.2<z<1.8$}\\
\noalign{\smallskip}\hline\noalign{\smallskip}
M$>1\times 10^{13} \ M_{\odot}$         &   95.4\% &  4.6\% &  76.4\% & 5.7\%\\
M$>2 \times 10^{13} \ M_{\odot}$ &   95.3\% &  4.6\% &  84.3\% & 0\%\\
M$>3 \times  10^{13} \ M_{\odot}$&   100\%  &    0\% &  82.4\% & 2.9\%\\
\noalign{\smallskip}\hline\noalign{\smallskip}
& \multicolumn{4}{c}{$1.8<z<2.5$}\\
\noalign{\smallskip}\hline\noalign{\smallskip}
M$>1\times 10^{13} \ M_{\odot}$         & 76\% &  3\% &  33\% & 0\%\\
M$>2 \times 10^{13} \ M_{\odot}$ & 78\% &  0\% &  39\% & 0\%\\
M$>3 \times  10^{13} \ M_{\odot}$& 75\% &  0\% &  37\% & 0\%\\
\noalign{\smallskip} \hline
\end{tabular}
\end{center}
\end{table}

\begin{table}[htb]
\caption{Average distances of detected peaks from real centres.}
\label{tab:simulazioni1a}
\begin{center}
\begin{tabular}{ccc}
\hline \noalign{\smallskip} \hline\noalign{\smallskip}
Redshift Interval & $\Delta r$ (Mpc)  &  $\Delta z$   \\
\noalign{\smallskip}\hline\noalign{\smallskip}
$0.4<z<1.2$ &  0.13 $\pm$ 0.09 &  0.016 $\pm$ 0.013 \\
$1.2<z<1.8$& 0.13 $\pm$ 0.07 &  0.028 $\pm$0.019 \\
$1.8<z<2.5$& 0.20 $\pm$ 0.11&  0.044 $\pm$0.033     \\
\noalign{\smallskip} \hline
\end{tabular}
\end{center}
\end{table}

\begin{table}[htb]
\caption{Separation threshold for aligned groups.}
\label{tab:simulazioni2}
\begin{center}
\begin{tabular}{cc|rr}
\hline \noalign{\smallskip} \hline\noalign{\smallskip}
 \multicolumn{2}{c}{Distance} & \multicolumn{2}{|c}{Separation threshold} \\
\hline\noalign{\smallskip}
 Redshift  & Projected  & $z \sim 1$   & $z \sim 2$ \\
\noalign{\smallskip}\hline\noalign{\smallskip}
  &$\Delta r$=1.0 Mpc &     $>8\sigma$      &  $>8\sigma$      \\
 1$\sigma_z$ &$\Delta r$=1.5 Mpc & 6$\sigma$    & 5$\sigma$\\
  &$\Delta r$=2.0 Mpc & 4$\sigma$    & 5$\sigma$\\
\noalign{\smallskip}\hline\noalign{\smallskip}
 & $\Delta r$=1.0 Mpc &    $>8\sigma$    &   $>8\sigma$      \\
 2$\sigma_z$& $\Delta r$=1.5 Mpc & 4$\sigma$    & 5$\sigma$\\
  &$\Delta r$=2.0 Mpc & 4$\sigma$    & 5$\sigma$\\
\noalign{\smallskip}\hline\noalign{\smallskip}
& $\Delta r$=1.0 Mpc & 5$\sigma$   & 5$\sigma$\\
 3$\sigma_z$& $\Delta r$=1.5 Mpc & 4$\sigma$    & 5$\sigma$\\
 & $\Delta r$=2.0 Mpc & 4$\sigma$    & 5$\sigma$\\
\noalign{\smallskip} \hline
\end{tabular}
\end{center}
\end{table}
Finally, we create different mock catalogues superimposing a number
of structures on the random fields. Given the relatively small
comoving volume sampled by the survey, we  expect to find only
groups  and small clusters with a total mass $M\sim \
10^{13}-10^{14} \ M_{\odot}$ and a number of members corresponding
to the lowest Abell richness classes \citep{girardi}. To check that
the performance of the algorithm does not change appreciably with a
varying number of real overdensities of this kind, we perform three different
subset of simulations. Each subset is based on the analysis of 10 mock catalogues, with a number of
groups equal to the number  of $M> 10^{13} \ M_{\odot}$, $M>2 \times
10^{13} M_{\odot}$ and $M>3 \times 10^{13} M_{\odot}$ DM haloes,
obtained by integrating the Press $\&$ Schechter function
\citep{press1974} over the comoving volume sampled by the survey.
Their positions in real space are chosen randomly. Cluster galaxies
follow a King-like spatial distribution $n(r) \propto
[1+(r/r_c)^2]^{-3/2}$  \citep[see ][]{sarazin} with a typical core
radius $r_{c} = 0.25  Mpc$. 

To take into account the uncertainty on photometric redshifts,  to each cluster we
assign  galaxy a random redshift extracted from a
gaussian distribution centred on the cluster redshift $z_{cl}$ and
having a dispersion $\sigma_z=0.03 \cdot (1+z_{cl})$. We neglect the
cluster real velocity dispersion, which is much smaller than the
$z_{phot}$  uncertainty. We analyse the simulations in the same way
as the real catalogue, i.e. calculating galaxy volume density
considering objects with $M_B \leq -18$ at $z< 1.8$ and objects with
$M_B \leq -19$ at $z \geq 1.8$.

We evaluate the \textit{completeness} of the sample of detected
clusters (fraction of real clusters detected) and its
\textit{purity}  (fraction of detected structures corresponding to
real ones) at different redshifts (see Table \ref{tab:simulazioni1}). We present also the number of
unresolved pairs (a detected structure corresponding to two real
ones) and the number of double identifications (a unique real
structure separated in two detected ones).

Our aim is to study the properties of individual structures and not, for example, to perform group number counts for
cosmological purposes. For this reason,  we prefer to choose conservative selection
criteria in order to maximise the purity of our sample, while still
keeping the completeness high. We isolate the structures  as described in Sect. \ref{algorithm}, and we consider as significant only those overdensities with at least 5 members in the $4\sigma$ region and 15 members in the $2\sigma$ region. 

A structure in the input catalogue is identified if its center is within $\Delta r=0.5 Mpc$ projected
distance, and within $\Delta z=0.1$, from the center of a detected
structure, for the low redshift sample and $\Delta r=0.8 Mpc$,
$\Delta z=0.2$ at high \textit{z} (to account for the increased uncertainties in
redshift  and position). The results are reported in Table
\ref{tab:simulazioni1}. We can see that the chosen thresholds and
selection criteria allow for a high \textit{purity} ($\sim 100 \%$)
at $z <1.8$, still detecting about the 80\% of the
real structures. At $z >1.8$, given the greatly
reduced fraction of observed galaxies, the noise is higher and these
criteria turn out to be very conservative (therefore the
\textit{completeness} is low) but are necessary to keep a low number
of false detections (\textit{purity} $\sim 75-80 \%$).  Table \ref{tab:simulazioni1a} shows the average distance between the
centres of the real structures and the centres of their detected
counterparts. The density peaks allow to identify the positions of real groups with a
good accuracy.

We also evaluate the ability of the algorithm to separate real
structures that are very close both in redshift and angular
position. In Table \ref{tab:simulazioni2} we present, for different
intracluster distances, the density level at which couples of real
groups appear as separated peaks. Both at low and high redshift it
is not possible to separate structures whose centres are closer than
1.0 Mpc on the plane of the sky and 2$\sigma_z$ in redshift. For
larger separations, using higher thresholds (5 or 6  $\sigma$ above
the average $\rho$) it is possible to separate the groups.

\section{A catalogue of the detected overdensities in the GOODS-South field}\label{sec:overdens}

\begin{sidewaystable*}
\caption{Overdensities in the GOODS-South field.} \label{tab:catalog}
\begin{center}
\begin{tabular}{lccccccccccccccc}
\noalign{\smallskip} \hline \noalign{\smallskip}
\hline\noalign{\smallskip} 
ID &Redshift & RA           & DEC     &  Members              &  Field & M$_{200}$     b=2-1&r$_{200}$ b=2-1& Peak Overdensity    &  
Numb. &   Spectroscopic   &    $\sigma_{v}$              & $r_{vir}$       & $M_{vir}$           \\     
   &         & (J2000)        & (J2000)   &                       &  &     ($ M_\odot/10^{14}$)              & ($Mpc$)            & $\sigma$   &     of $z_{spe}$                     &     position                &     $(km\ s^{-1})$
   & (Mpc)          & ( $M_\odot/10^{14}$)     \\  
\noalign{\smallskip}\hline\noalign{\smallskip}
$1$          &0.66  &    53.1623   &  -27.7913    &  19  & 6  & 0.15- 0.3& 0.9-1.1 &$7 $ &6 & $0.665\pm0.001$&$446\pm180$  & 0.85 & 1.0   \\
$2^{a} $     &0.66  &    53.0630   &  -27.8280    &  50  & 17 & 0.2-0.4& 1.1-1.3 &$10 $  &  & & &                                         \\
$3  $        &0.69  &    53.1690   &  -27.8747    &  54  & 20 & 0.3 - 0.5& 1.1-1.4 &$6 $ &  & & &                                         \\
$4^{a,b,c}$  &0.71  &    53.0797   &  -27.7920    &  92  & 37 & 0.9 - 3.0& 1.7-2.4 &$10$  &36& $0.734\pm0.001$ &$634\pm107$  & 1.3  & 3.2  \\
$5  $        &0.96  &    53.0843   &  -27.9020    &  32  & 14 & *        &  *       &$6$  &  & & &                                         \\
$6^{c}$      &1.04  &    53.0570   &  -27.7693    &  57  & 31 & 0.5 - 1.1& 1.4-1.8 &$8 $ &  & & &                                           \\
$7$          &1.04  &    53.1577   &  -27.7660    &  60  & 26 & 0.4 - 0.8& 1.2-1.6 &$6 $ &  & & &                                           \\
$8^{b,d}$    &1.06  &    53.0697   &  -27.8773    &  38  & 18 & 0.2 - 0.5& 1.1-1.3 &$10$  &6 &$1.0974\pm 0.0015$  & $446\pm 143$ &0.8 & 0.8 \\
$9^e $       &1.61  &    53.1270   &  -27.7140    &  50  & 24 & 2.0 - 4.9& 2.1-2.9 &$7 $ &6 &$1.610\pm 0.001$  &$482\pm217$ &1.4 &1.46      \\
$10$         &2.23  &    53.0763   &  -27.7060    &  20  &  8 & 0.8 - 1.4& 1.5-1.9 &$6 $ &  & & &                                           \\
$11$         &2.28  &    53.1470   &  -27.7087    &  23  & 12 & 0.6 - 1.3& 1.5-1.8 &$10$  &  & & &                                           \\
$12$         &2.28  &    53.0970   &  -27.7640    &  19  & 7  & 0.6 - 1.6& 1.3-2.0 &$9 $ &  & & &                                           \\
\noalign{\smallskip} \hline
		\end{tabular}
\\
\medskip
\begin{tabular}{l}
a - \cite{gilli2003}\\
b - \cite{adami2005} \\
c - \cite{trevese2006}\\
d - \cite{diaz2007}\\
e - \cite{castelllano2007}\\
\end{tabular}
\\
\scriptsize{* We do not present $M_{200}$ and $r_{200}$
for structure 5 because it is located on the edge of the field and
it is very close to structure 8, see fig \ref{fig:strutture}.}
\end{center}
\end{sidewaystable*}

\begin{figure}[htb]
\resizebox{\hsize}{!}{\includegraphics{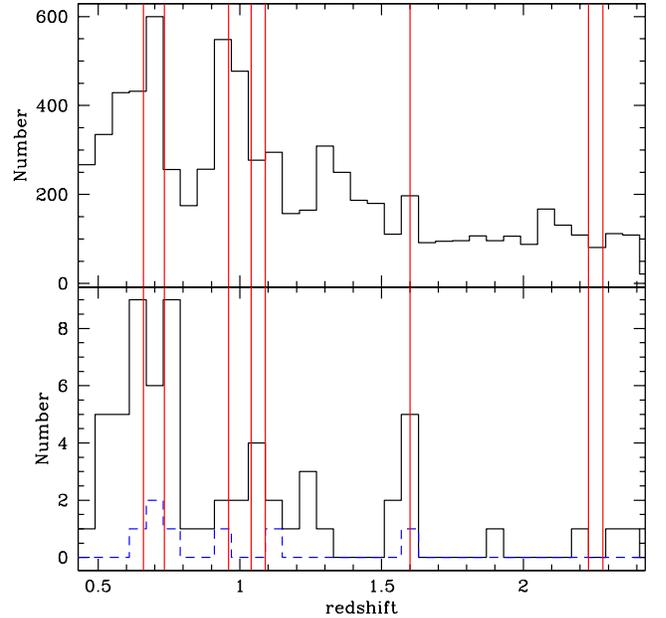}} \caption{Upper panel: photometric redshift  distribution of our sample (continuous line). Vertical lines mark the redshifts of the detected
structures. Lower panel: redshift distribution  of spectroscopically selected AGNs in
the GOODS-South field (continuous line); the dashed-line histogram is the distribution of
the AGNs associated with the overdensity peaks in Table
\ref{tab:catalog}.} \label{fig:agn}
\end{figure}

\begin{figure*}[!htb]
\resizebox{\hsize}{!}{\includegraphics{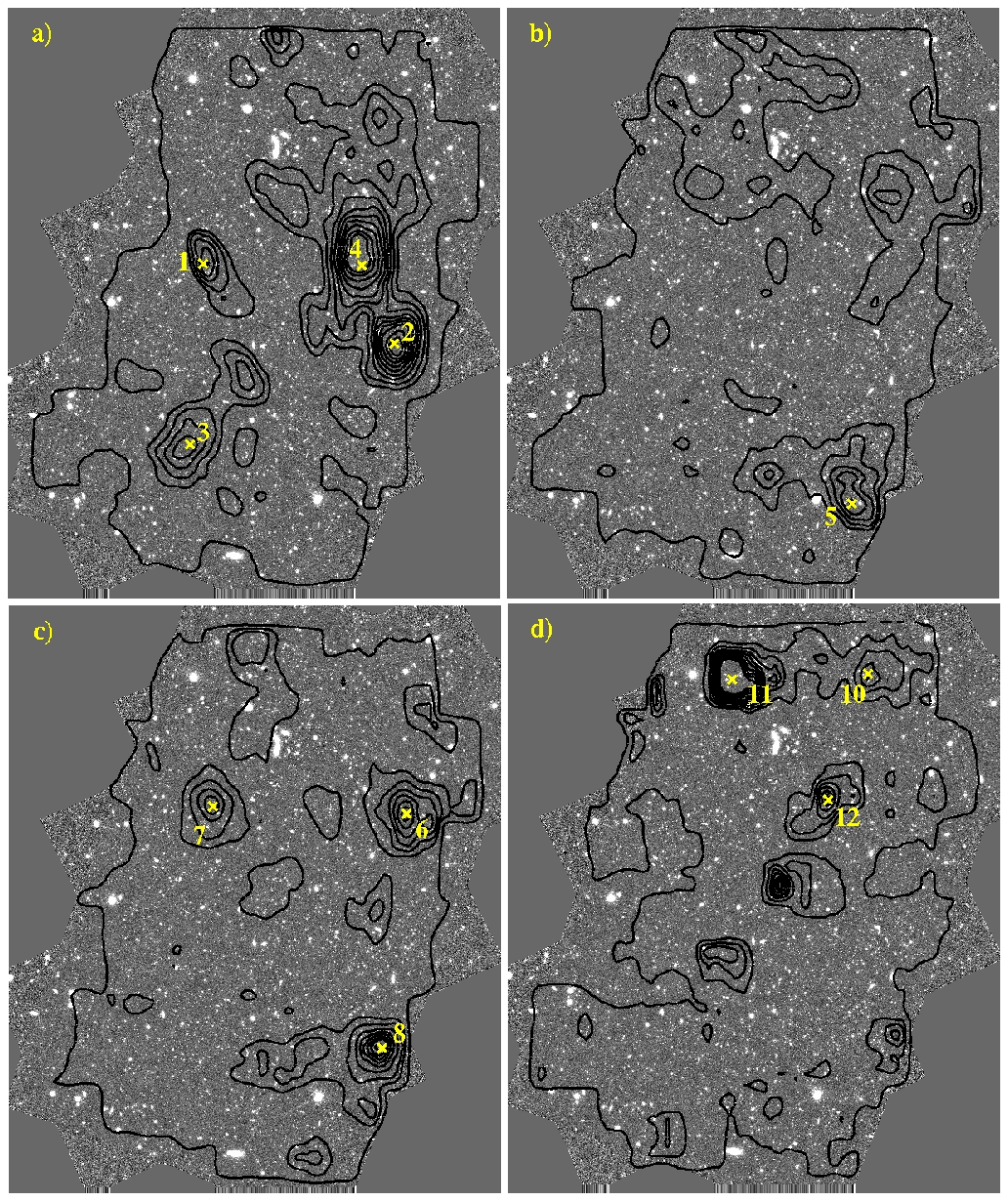}} \caption{Density isosurfaces for structures at  $z\sim 0.7$ (a), at
$z\sim 0.95$ (b), $z\sim 1.05$ (c)   and at $z\sim 2.3$ (d)  (average, average $+2\sigma$, average $+3\sigma$ to average $+10 \sigma$)
superimposed on the ACS $z_{850}$ band image of the GOODS-South field. Yellow crosses indicate the density peak of each structure, the number is the ID of the structure in Table\ref{tab:catalog}. For the analogous picture regarding the cluster ID=9 at $z\sim 1.6$ see \citet{castelllano2007}. Other overdensities present  did not pass our selection criteria described in Sect. \ref{simulazioni}.} \label{fig:strutture}
\end{figure*}

An  inspection of the 3-D density map shows some complex high
density structures distributed over the entire GOODS field. In
particular, we find diffuse overdensities at $z\sim 0.7$, at
$z\sim 1$,  at $z\sim 1.6$  and at $z\sim 2.3$. Some of these have
already been partially described in literature
\citep{gilli2003,adami2005,vanzella2005,
trevese2006,diaz2007,castelllano2007}.  Fig. \ref{fig:agn} shows the position of these overdensities  over the photometric redshift distributions of our sample.These overdensities are also traced by the distribution of the spectroscopically confirmed AGNs in our catalogue, as shown in the lower panel of Fig. \ref{fig:agn} (these objects are not included in the sample used for the density estimation). This link between large scale structures and AGN distribution was already noted, at lower redshift, in the CDFS
\citep{gilli2003}, in the E-CDFS \citep{silverman2007} and in the CDFN  \citep{barger2003}.

Within these large scale overdensities, we identify the
structures, with the procedure described in Sect. \ref{simulazioni}.
Using an analysis with a  $5\sigma$ threshold, we find that two structures identified with $\rho>\bar{\rho}+4\sigma$,  at $z \sim
0.7$ and $z \sim 1$,    are the sum of two different structures, so
we used a $5 \sigma $ threshold to separate these peaks. We then
associate the galaxies belonging to the region of overlap between the two structures to the
less distant peak. 

Overall, we find four structures at $z\sim
0.7$, four structures at $z\sim 1$, one at $z \sim 1.6$ (see also
C07) and three structures at $z\sim 2.3$. The density isosurfaces of  the structures at  $z\sim 0.7$, at
$z\sim 0.96$, $z\sim 1.05$  and at $z\sim 2.3$  are shown in Fig. \ref{fig:strutture},
superimposed on the ACS $z_{850}$ band image of the GOODS-South. The analogous image for the  overdensity at  $z\sim 1.6$ is showed in \cite{castelllano2007}.  In the figure, we indicate with a cross the peak position of  the identified structures. Other overdensities present  did not pass our selection criteria described in Sect. \ref{simulazioni}.

All the structures are presented in Table \ref{tab:catalog}, where
we list the following properties:\\
{\it Column 1}: ID number. \\
{\it Column 2-4}: The position of the density peak (redshift, RA and DEC)
obtained with our 3-D photometric analysis.\\
{\it Column 5}: The number of the objects associated with each structure
as defined above. This number gives a hint on the richness of the
structure; however it should not be used to compare structures
at different
redshifts because of the different magnitude intervals sampled.\\
{\it Column 6}: The average number of field objects present in a volume
equal to that associated to the structure, at the relevant redshift. We
calculated this number by integrating the evolutive LFs obtained by
\cite{salimbeni2008}. In particular, we integrated the LF up to an
absolute limiting magnitude calculated using the average K- and
evolutionary corrections and $z_{850}$ limiting observed magnitude
as done in Sect. \ref{simulazioni}. In this way we take into account
the selection effects given by the magnitude cut in our catalogue,
as a function of redshift.\\
{\it Column 7-8}: The $M_{200}$  and $r_{200}$ (assuming bias factors 1
and 2). The mass $M_{200}$ is defined as the mass inside the radius
corresponding to a density contrast $\delta_{m}=\delta_{gal}/b$
$\sim$ 200 \citep{carlberg}, where $b$ is the bias factor. To estimate the 3D galaxy density
contrast $\delta_{gal}$ we count the objects in the photometric
redshift range occupied by the structure as a function of the
cluster-centric radius. We then perform a statistical subtraction
of the background/foreground field galaxies, using an area at least
2.5 Mpc (comoving) away from the center of every cluster  in the
relevant redshift interval. Finally, the density contrast is
computed assuming spherical symmetry of the structure.
The mass inside a volume V of density contrast $\delta_{gal}$ is
determined adapting to our case the method used for spectroscopic
data at higher $z$ by \citet{steidel}:
\begin{equation}\label{eq:steidel} M=\bar{\rho_{u}} \cdot V \cdot (1+\delta_{m}),
\end{equation} in which $\bar{\rho_{u}}$ is the average density of the Universe and $\delta_{m}$ is the total mass density contrast related to the galaxy number density contrast through a bias factor: $1+\delta_{m}=1+\delta_{gal}/b$. We assume a
  bias factor $b$ in the range $1\le b \le 2$ \citep[see][]{arno}.\\
{\it Column 9}: The level of the density peak, measured in
number of $\sigma$ above the average volume density.

We then searched the available spectroscopic public data
\citep{wolf2001,lefevre2004,szokoly2004,mignoli2005,vanzella2005,vanzella2006,vanzella2007}
to check if any of the members of the structures have spectroscopic
redshifts, to estimate their location and the velocity dispersion
when the statistics is sufficient. Spectroscopic galaxies are
considered members of a structure if their redshift was within $4500
km/s$ (i.e. three times the velocity dispersion of a rich cluster)
from the mode of its redshift distribution. From these data, we
estimated the average redshift of the structure and the velocity
dispersion using the biweight estimators, computed using the Rostat
package \citep{beers1990}, with $68\%$ confidence uncertainties
obtained from a Jackknife analysis.

In Table $\ref{tab:xray}$ we present a value for the X-ray count rate in the band 0.3-4 kev,
the corresponding flux (in the interval 0.5-2 keV) and the
rest-frame luminosity (0.1-2.4 keV), from the Chandra 2Ms exposure
\citep{luo2008}. We measure the count rates in a square of
side of $\sim 30 \arcsec$, centred on the position of the peak of
each structure. For the count-rate to flux conversion we assume as spectrum a Raymond-Smith model \citep{raymond1977} with T=1 keV and 3 keV and metallicity of 0.2 $Z_\odot$.

\begin{table}[htb]
\caption{X-ray observations.} \label{tab:xray}
\begin{center}
\begin{tabular}{ccccc}
\noalign{\smallskip}\hline\noalign{\smallskip}\hline\noalign{\smallskip}
ID& Count Rate& Flux$^a$                                & ${L_X}^{a}$        & S/N$^{b}$\\
  & 0.3-4 keV& 0.5-2 keV                       &0.1-2.4 keV   &    \\
  & ($10^{-5}$)  & ($10^{-16} \ erg \ s^{-1} \ cm^{-2}$) &($ 10^{43}\ erg \ s^{-1}$) &     \\
\noalign{\smallskip}\hline\noalign{\smallskip}
1  & 8.49 &6.80-9.01&0.12-  0.26& u.l.\\
2  & 5.56 &4.45-5.90&0.08-  0.18& u.l.\\
3  & 10.1 &8.15-10.98&0.16-  0.37& u.l.\\
4  & 11.2 &9.04-12.31&0.19-  0.44& u.l.\\
5  & 23.7 &19.31-29.21&0.86-  2.36& 11.3 \\
6  & 5.90 &3.04-4.14&0.26-  0.76& u.l.\\
7  & 5.77 &2.97-4.05&0.26-  0.74& u.l.\\
8  & 9.88 &5.10-6.91&0.47-  1.37& u.l.\\
9  & 5.68 &3.08-4.14&0.83-  3.67& u.l.\\
10 & 9.37 &5.39-7.54&3.50- 22.43& u.l.\\
11 & 5.72 &3.29-4.69&2.27- 15.06& u.l.\\
12 & 6.70 &3.85-5.50&2.66- 17.64& u.l.\\
\noalign{\smallskip} \hline
\end{tabular}
\\
\medskip
\begin{tabular}{l}
a -  Values for a Raymond-Smith model with assumed temperature\\
  respectevelly of 3 keV and 1 keV and metallicity 0.2 $Z_\odot$.\\
b - u.l. indicates structures with a 3 $\sigma$ upper limit in the flux. 
\end{tabular}
\\
\end{center}
\end{table}

\subsection{Structures at $z\sim 0.7$}

At redshift $z\sim 0.67$ we isolate three high density
peaks (ID=1,2 and 3) that are part of a large scale structure
already noted, as a whole, by \cite{gilli2003}.

For the structure with ID=1, we estimate the redshift from the
available 6 spectroscopic data. We find an average redshift of
$0.665\pm 0.001$ and a velocity dispersion of $446\pm 180 \ km \
s^{-1}$. Assuming that the cluster is virialised, we estimate
$r_{vir}=0.8 Mpc$ and  $M_{vir}=1.0 \cdot 10^{14} M_\odot$, using
the relations in \cite{girardi1998}. This estimate is based also
on the assumption that there
are no infalling galaxies and that the surface term \citep[e. g.
][]{carlberg1996} is negligible. Considering the uncertainties, also
due to the small number of spectroscopic galaxies, $M_{vir}$ is
fairly consistent with the $M_{200}$ estimated from the galaxy
density contrast ($0.9-3 \cdot 10^{14} M_\odot$).

We also derive the upper limits on the X-ray luminosities for this
structure, that is of the order of $0.2-0.3 \cdot 10^{43} \ erg \
s^{-1}$.  All the properties presented are
consistent with the structure being a galaxy group/small cluster
\citep{bahcall1999}.

The structures with ID=2, 3 have upper limits on their X-ray
luminosities of the order of $0.2-0.3 \cdot 10^{43} \ erg \ s^{-1}$,
and their masses are of the order of $M_{200}\sim 0.2-0.5 \cdot
10^{14} M_\odot$. These X-ray luminosities and masses are all
typical of galaxy groups/small clusters \citep{bahcall1999}. Each of these structures contains a spectroscopically confirmed galaxy  detected in the VLA 1.4 GHz survey \citep{miller2008}. 

At a slightly  higher redshift ($z\sim 0.7$) we identify a
 high density peak (ID = 4) embedded in another large scale
structure which was already known in literature
\citep{gilli2003,adami2005,trevese2006}. In our previous paper
\citep{trevese2006}  we identify this structure applying our
algorithm to the data from the K20 catalogue, and classified it as
an Abell 0 cluster. 

In this new analysis we find that this
structure is symmetric and has a regular mass profile. It has 92 associated objects ($M_B(AB)<-18 $),
and two AGNs. From the density contrast we obtain an
$r_{200}=1.7-2.4 Mpc$ and a total mass of $M_{200}=0.9-3.0 \cdot
10^{14} M_{\odot}$ for bias factor b=2-1.  From the 36 galaxies with spectroscopic redshifts, we estimate a redshift location of
$0.734\pm 0.001$ and a velocity dispersion of $634\pm 107 km \
s^{-1}$. We derive a virial radius $r_{vir}=1.3 Mpc$, and a virial
mass $M_{vir}=3.2 \cdot 10^{14} M_\odot$, in good agreement
with $M_{200}$. The 3 sigma upper limit for the X-ray luminosity in
the interval 0.1-2.4 keV is very low ($L_X =0.19-0.44 \cdot 10^{43} \ erg
\ s^{-1}$). Note that the area we considered does not include
the X-ray source 173 of \cite{luo2008}, that, similarly to
\cite{gilli2003}, we associate to the halo of the brightest cluster
galaxy (ID$_{GOODS-MUSIC}$=9792). Alternatively, \cite{adami2005}
associated the bolometric luminosity ($L_X=0.11 \cdot 10^{43} \ erg
\ s^{-1}$) of the X-ray source 173 to the thermal emission of the
intra-cluster medium (ICM). From this value they deduced a galaxy
velocity dispersion around $200-300 \ km \ s^{-1}$. This value is
apparently in contrast with the $\sigma_v$ estimated from the
spectroscopic redshifts. We also associate to the galaxy ID$_{GOODS-MUSIC}$=9792 the object 236 detected in the VLA 1.4 GHz survey. It  has an integrated emission of $517.5\pm13.1\mu$ Jy  \citep{miller2008}.

\begin{figure}[htb]
\resizebox{\hsize}{!}{\includegraphics{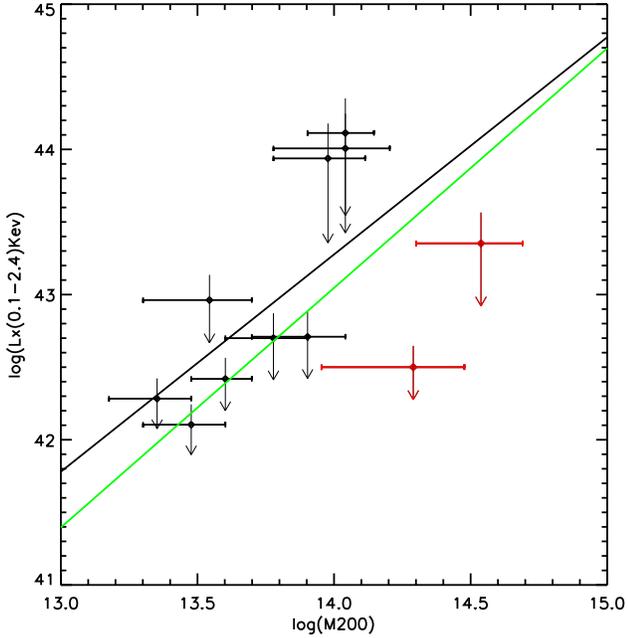}} \caption{L$_X$ vs M$_{200}$ for the clusters in Tab \ref{tab:xray}. The horizontal error bar is calculated considering a bias factor in the range $1\le b \le 2$, while the vertical error bars are computed varying the gas temperature between T=1 keV and T=3 keV as discussed in the main body. 
The clusters at $z\sim 0.7$ and $z\sim 1.6$ are indicated by red points and error bars. The $M_{200}-L_{X}$ relations found by \cite{reiprich2002}
and by \cite{rykoff2008} are indicated by a black and green line respectively. } \label{fig:x}
\end{figure}

From this analysis we can conclude that our two independent mass
estimates ($M_{200}$ and $M_{vir}$) are consistent with this
structure being a virialised poor cluster. However, the X-ray
emission is significantly lower than what is expected from its
optical properties, as it shows from the comparison in Fig. \ref{fig:x} with the M$_{200}$-L$_X$ relations found by \cite{reiprich2002} and by \cite{rykoff2008}.

\subsection{Structures at $z\sim 1$}

At redshift $\sim 1$ we find four structures (ID= 5, 6, 7 and  8).

The structure with ID=5  at $z\sim 0.96$ has 32 member
galaxies. This structure can be associated to the
extended X-ray source number 183 in the catalogue by
\cite{luo2008} derived from the 2MS Chandra observation. This extended
X-ray source had not been associated to any structure so far. From the count rate
in the interval 0.3-4 keV (S/N=11.3) we estimate a luminosity
$L_X=0.86-2.36 \cdot 10^{43} \ erg \ s^{-1}$ (in the interval 0.1-2.4 keV). 
For the structures with ID=6, 7 we estimate $r_{200}\sim
1.2-1.8 Mpc$, and a total mass of $M_{200}=0.4-1.1 \cdot 10^{14}
M_{\odot}$.  The 3 sigma upper limits for their X-ray luminosity are
all slightly below $10^{43} \ erg \ s^{-1}$, consistent with their $M_{200}$ masses. 

The structure with ID=8 at $z\sim 1.06$, has 38 associated galaxies,
and an AGN spectroscopically confirmed. We derive a precise
redshift location of $z=1.0974\pm0.0015 $ and a velocity dispersion of
$446\pm 143 km sec^{-1}$, from 6 galaxies with spectroscopic
redshift. From these galaxies we also obtain $M_{vir}=0.8 \cdot
10^{14} M_\odot$ and $r_{vir}=0.8 Mpc$. We estimate
$r_{200}=1.1-1.3 Mpc$, and $M_{200}=0.2-0.5 \cdot 10^{14}
M_{\odot}$, which are compatible values with a group of such  $M_{vir}$ and $r_{vir}$. This structure was already found with different
methods by \cite{adami2005}, using a friend-of-friend algorithm on
spectroscopic data from the VIMOS VLT survey (structure 15 in their
Table 4), and by \cite{diaz2007} studying the extremely red objects
on GOODS-South (they call this structure GCL J0332.2-2752). Their
redshift positions and the velocity dispersions are consistent with
those obtained in the present analysis. The 3 sigma upper limit for
the X-ray luminosity is around $ 10^{43} \ erg \ s^{-1}$,
consistently with the estimated $M_{200}$ mass.

Considering their properties, these four structures can be
classified as groups of galaxies. Consistent results for the
structure with ID=6 were obtained in \cite{trevese2006}.

\subsection{Structures at high $z$}

At redshift $z\sim1.6$, we find a compact structure that corresponds to a forming cluster, as
already discussed  in detail by C07  \citep[see also][]{kurk2008}. We find a
regular mass profile for this structure, and we estimate an
$r_{200}=2.1-2.9 Mpc$, and a $M_{200}=2.0-4.9 \cdot 10^{14}
M_\odot$. This structure has 50 members,
including 3 spectroscopic redshifts, and a confirmed AGN, from
the GOODS-MUSIC catalogue. We add three other spectroscopic
redshifts from the GMASS sample \citep{cimatti2008}. From these 6
redshifts we estimated a velocity dispersion of $482\pm 217 km/s$,
and derived an $M_{vir}=1.4 \cdot 10^{14} M_\odot$ and $r_{vir}=1.46
Mpc$. This estimate is consistent with the value in Table
\ref{tab:catalog}. We derive an upper limit to the X-ray luminosity of
$0.83-3.67 \cdot 10^{43} \ erg \ s^{-1}$ (0.1-2.4 KeV), lower than expected from the
velocity dispersion and the estimated $M_{200}$ (see Fig. \ref{fig:x}).

At $z\sim 2.2$  we find a diffuse overdensity, similar to those at
lower redshift, embedding three structures. We associate to these
structures 20, 23 and 19 galaxies.
We estimate for all these structure an $r_{200}\sim 1.3-2 Mpc$ and a mass of $M_{200}\sim
0.6-1.6 \cdot 10^{14}\ M_\odot$. These structures appear to be
comparable to those at $\sim 0.7$ and $\sim 1.6$, and they could be
forming clusters.

\section{Colour-Magnitude diagrams}

\begin{figure}[!ht]
\resizebox{\hsize}{!}{\includegraphics{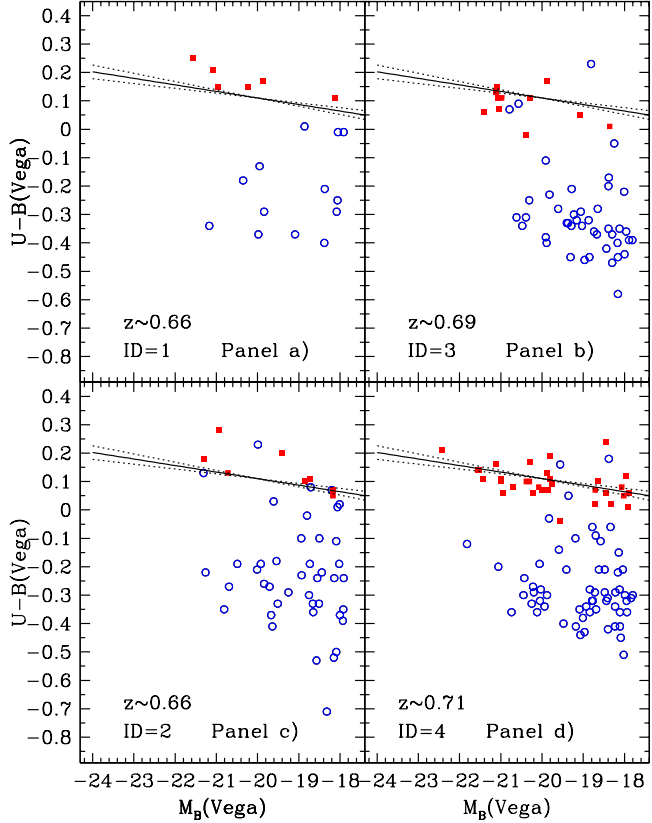}} \caption{Rest frame
colour magnitude relations ($U-B$ vs $M_B$) for each structure at
$z\sim 0.7$. Squares indicate passively evolving galaxies selected as
$age/\tau\ge 4$, and the circles are galaxies with $age/\tau < 4$.
Filled points indicate galaxies with spectroscopic redshift. The
continuous lines are the fit to the {\it red sequence} of all the
combined structures. The dotted lines are the uncertainties at 1-sigma obtained
with a Jackknife analysis.}\label{fig:colmag}
\end{figure}

\begin{figure}[!ht]
\resizebox{\hsize}{!}{\includegraphics{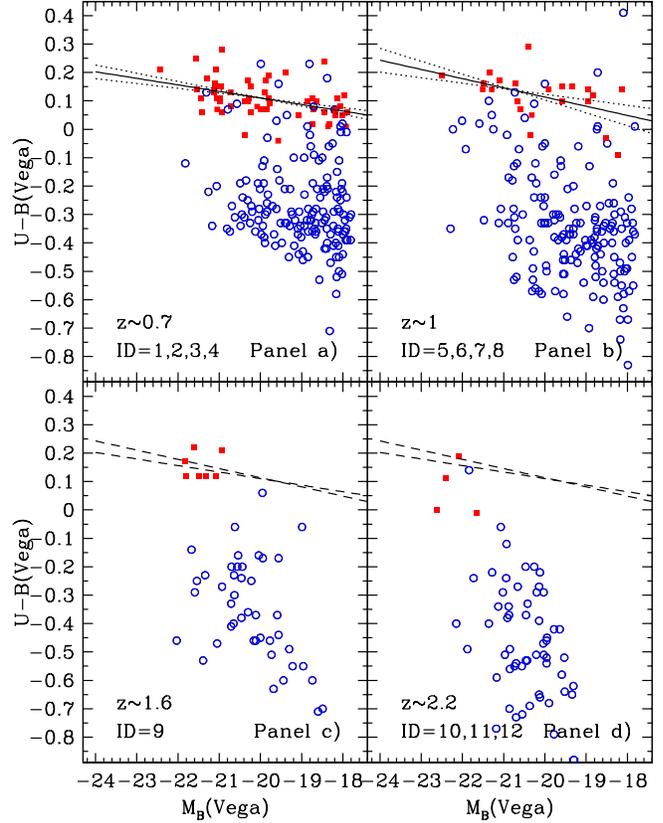}} \caption{Same as
Fig. \ref{fig:colmag}. Panel a: structures at $z\sim 0.7$; panel b:
structures at $z\sim 1$; panel c: structure at $z\sim 1.6$ ; panel
d:  structures at $z\sim 2.2$. The two dashed lines in each of the two last bins
of redshift are the red sequences estimated at $z\sim 0.7$ (shallower slope) and $\sim
1$ (steeper slope)}\label{fig:colmag2}
\end{figure}

We study the colour magnitude diagrams ($U-B$ vs. $M_B$) for all
the structures, as shown in Figs. \ref{fig:colmag} and
\ref{fig:colmag2}. To estimate the slope of the red-sequence, we
define its members as passively evolving galaxies according to the
physical criterion $age/\tau\ge 4$, where the age and $\tau$ (the
star formation e-folding time) are inferred for each galaxy from the SED fitting (Sect. \ref{catalog}). This quantity is, in practice, the
inverse of the Scalo parameter \citep{scalo1986} and a ratio of 4 is chosen to
distinguish galaxies having prevalently evolved stellar populations
from galaxies with recent episodes of star formation. Indeed, an
$age/\tau=4$ corresponds to a residual  2\% of the initial SFR,
for an exponential star formation history, as adopted in this paper.
\cite{grazian2006_eros} showed that this value can be used to effectively
separate star forming galaxies from the passively evolving population
\citep[see][ also for the discussion on the uncertainty associated to  this
parameter]{grazian2006_eros}. Passively evolving galaxies are
indicated in figures as filled squares.

Fig. \ref{fig:colmag}  shows the colour magnitude diagrams for
the four structures between $z=0.66$ and $z=0.71$. The cluster at
$z\sim 0.71$ (Panel d) shows a well defined red sequence, while the
three structures at $z\sim 0.66$ have fewer passively evolving
galaxies. Therefore, in order to increase our statistics, we estimate the colour-magnitude slope combining all the four
structures in the interval $0.66<z<0.71$  (see Panel a in Fig.
\ref{fig:colmag2}). We obtain a value $-0.023\pm
0.006$ for the slope. The resulting colour-magnitude relation is plotted in all
panels in Fig. \ref{fig:colmag} and in the panel a in Fig.
\ref{fig:colmag2} as a continuous line. The dotted lines constrain
the error at 1-sigma obtained with a Jackknife analysis. It is possible to see
in Fig. \ref{fig:colmag} that this average colour magnitude relation is  roughly consistent with the position in the (U-B) vs B diagram of the galaxies belonging to each single structure.  We therefore apply the same method at higher redshift, i. e. we
estimate the slope of the red sequence by combining the different
structures at the same redshift.

Fig. \ref{fig:colmag2}, panel b,  shows the colour magnitude diagram
for the structures at $z\sim 1$. We find a slope of $-0.03\pm
0.01$.

Panel c in Fig. \ref{fig:colmag2} shows the colour magnitude diagram
for the structure at $z\sim 1.6$. In this case we have  galaxies
distributed on less than a magnitude range, that is insufficient to
estimate the slope of the ``red sequence''. However, if we plot the
two sequences obtained at lower redshift, we can see that the few
passively evolving galaxies are consistent with them.

Finally, at redshift $\sim 2$, we have only 4 passive objects from the combination of 3 structures and there is no evidence of a well defined red sequence. We note that the colours of these objects are generally bluer in comparison to the colour of the relations found at lower redshifts.

The values of the slopes of the structures at redshift $\sim 0.7$
and $\sim 1$ are consistent with those of previous determinations
\citep[e.g.][]{blakeslee2003,homeier2006,trevese2006}. We confirm that the observations indicate
no evolution up to redshift $\sim 1$.
This would imply that the
mass-metallicity relation that produces the red sequence
\citep{kodama1998} remains practically constant up to, at least, $z\sim 1$.

\begin{figure*}[thb]
\resizebox{\hsize}{!}{\includegraphics[angle=270]{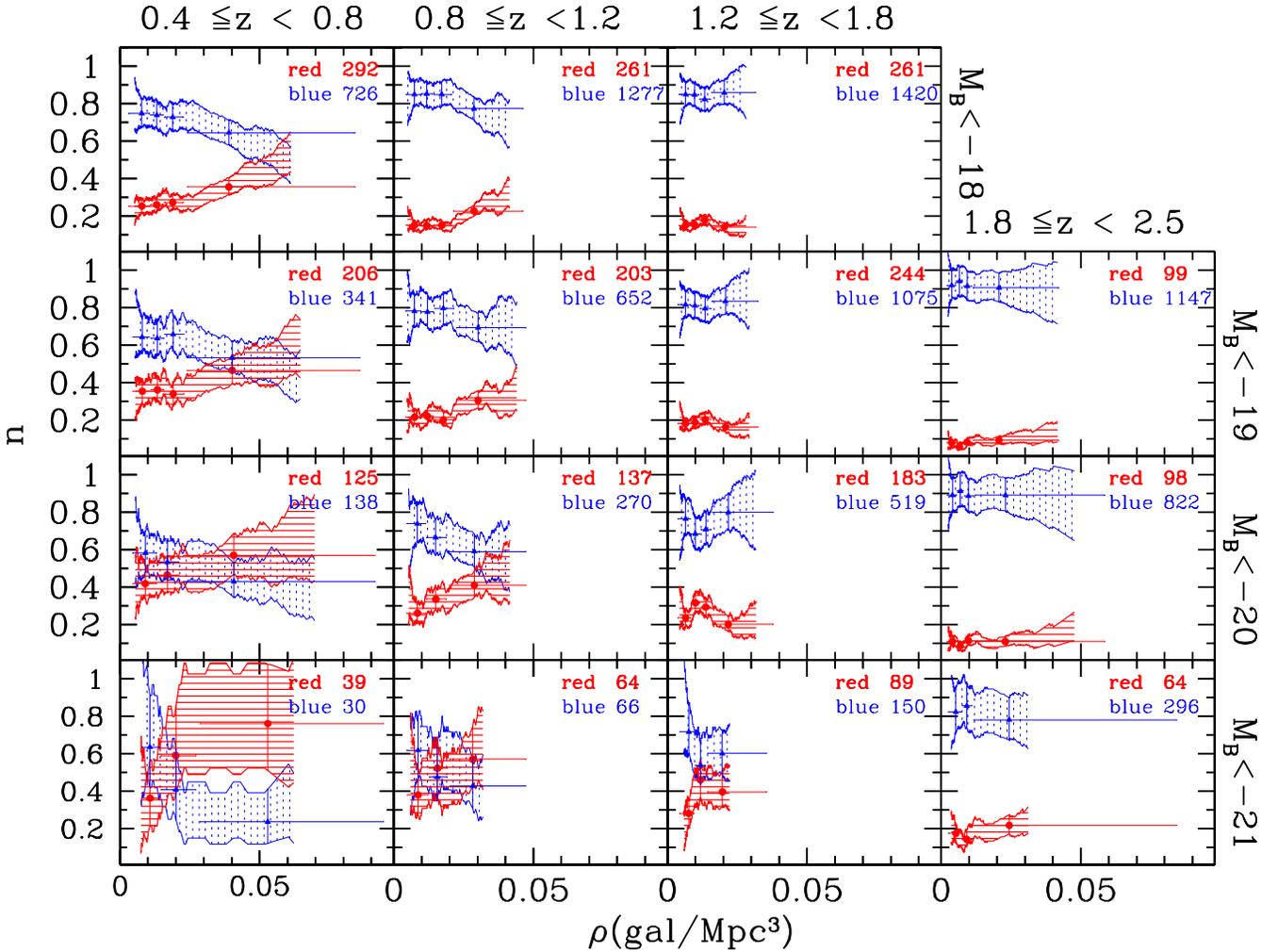}}
\caption{Fraction of red (filled circles) and blue galaxies (filled
triangles) at decreasing rest frame B magnitudes (from top to bottom) in four contiguous intervals of increasing redshift
(from left to right). Vertical errorbars indicate the poissonian uncertainty in
each bin. The shaded areas are obtained by smoothing the red (blue)
fraction with an adaptive sliding box. The horizontal errorbars
indicate the range of density covered by the 5-95 \% of the total
sample.} \label{fig:similcucciati}
\end{figure*}

\section{Galaxy properties as a function of the
environment}\label{sec:fieldprop}

To each object in the sample we associate the comoving density at
its position, and we study galaxy properties as a continuous
function of the environmental density.

\subsection{Galaxy populations: bimodality}
 We study the variation of the fraction of red and
blue galaxies as a function of the environmental density. To separate red and blue galaxies we use the minimum in the bimodal galaxy distribution in the (U-V) vs. B colour magnitude diagram, derived by
\cite{salimbeni2008}. Fig. \ref{fig:similcucciati} shows the
fraction of red and blue galaxies for different rest frame B
magnitudes in four redshift intervals. In general, for every
environment, we find that, at fixed luminosity, the red fraction
increases with decreasing redshift, and, at fixed redshift, it
increases at increasing B luminosity. We also find that for $z<1.2$
the red fraction increases with density for every luminosity, while
this effect is absent at higher redshift.

Our results extend to higher redshift those obtained by
\cite{cucciati06} on the VVDS survey, with a shallower spectroscopic
sample that reaches $z\sim 1.5$.  We find that at $z>1.2$ even
the highest luminosity galaxies are blue, star forming objects,  similarly to the results in  \cite{cucciati06},
although our colour selection is slightly different, since we select in colour two complementary samples, while they select two extreme red and blue populations ($(u^*-g')\ge 1.1$ and $(u^*-g')\le 0.55 )$).
Our results are also in agreement with the analysis of the DEEP2 survey
by \cite{cooper2007} in the redshift range $0.4<z<1.35$. They found
a weak correlation between red fraction and density at $z\sim 1.2$.
We see that at $z>1.2$ this correlation disappears, indicating that
the change probably occur in the critical range $1.5<z<2.0$, at
least in the environments probed by our sample. However we note
that, given the relatively small area covered, we do not probe very
high density regions  (i.e. rich clusters), at variance with wide,
low redshift surveys. When rich clusters are considered 
\citep[e.g. ][]{balogh2004}, a stronger variation with environment  in the
colours of faint galaxies is seen.  In any case, the disappearance at $z>1.2$ of the variation of the red fraction in the density range probed by our sample,
is an indication that a relevant change in
galaxy properties takes place at $z\sim 1.5 - 2$.

\subsection{Galaxy physical properties in high and low density environments}
We then study the distribution of physical parameters and photometric
properties for galaxies in high density environments, and
compare it to field galaxies. The first sample is
defined as the combination of the data from structures with similar
redshifts ('group galaxies' hereafter). The field galaxies are
defined as those with an associated $\rho$ lower than the median
density  (0.0126 for $z<1.8$  and 0.0085 for $z>1.8$)  of the entire sample ('field galaxies' hereafter). We
quantify the differences in the distributions of  the galaxy physical properties, i.e. mass, age, star formation rate, through the
probability $P_{KS}$  of the two samples, obtained as described above,  using a Kolmgorov-Smirnov test. We reject the
hypothesis that two samples  are drawn from the same distribution if
$P_{KS}< 5 \cdot 10^{-2}$. 

\begin{figure}[htb]
\resizebox{\hsize}{!}{\includegraphics{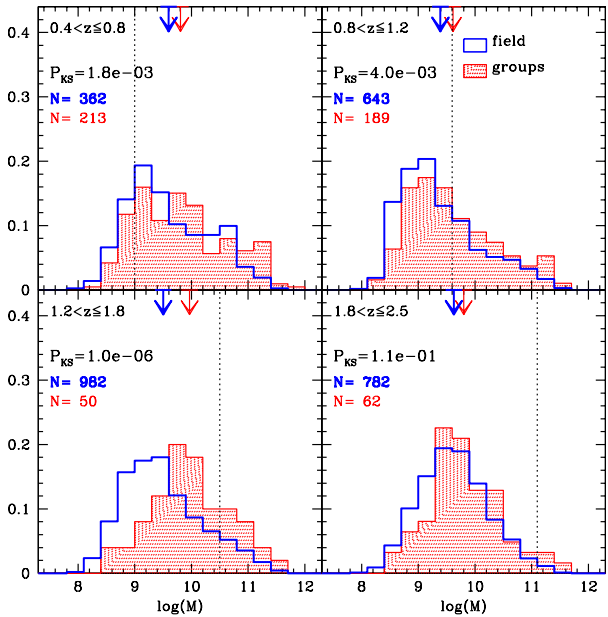}} \caption{Galaxy
stellar mass distribution in four redshift intervals. Shaded red
histograms represent galaxies associated with the density peaks and
empty black histograms represent galaxies in the low density regions, as
described in the text. In each panel the average value of $\log(M)$ for the two
distributions are indicated by arrows of the same colour. The K-S probability is
reported in each panel.  Vertical lines indicate the mass limit at the median redshift of the bin ($\log(M)=9.0$ at $z
\sim 0.6$, $\log(M)=9.6$ at $z \sim 1$,  $\log(M)=10.5$ at $z \sim 1.6$ and $\log(M) =11.1$ at $z \sim 2.15$).}\label{fig:mass}
\end{figure}

Fig. \ref{fig:mass} shows the distribution of the galaxy total
stellar mass in high and low density regions, in the same four
contiguous redshift intervals used before. The galaxies in high density
environment have a distribution that generally peaks at higher
masses with respect to ``field'' galaxies.  For the mass distribution we find a significant
difference in all but the last redshift bin as shown from the $P_{KS}$. It is important to
remark here that the shape of the distributions at low masses could
depend on the luminosity selection. In fact, a magnitude--limited
sample does not have a well defined limit in stellar mass. This
effect depends on the range of $M/L$ ratio spanned by galaxies with
different colours,  e. g. as shown in \cite{fontana06} in our sample, at $z\sim 1$,  $M/L_K$
extends from 0.9, for redder objects, to 0.046, for bluer objects. If a colour segregation is present as a function
of the environment, it could bias the distribution favouring the
observation of lower mass galaxies in less dense regions, where
the fraction of blue galaxies is higher. Although, as shown in Fig.
\ref{fig:similcucciati}, we do not find a strong colour
segregation, especially at $z>1$,  we carry out here also a more conservative analysis. 
We consider only the range of masses above the completeness mass limit obtained from the maximal $M/L_{z_{850}}$ for a passive evolving system ($\log(M)>9.0$ at $z
\sim 0.6$, $\log(M)> 9.6$ at $z \sim 1$,  $\log(M)> 10.5$ at $z \sim 1.6$ and $\log(M) > 11.1$ at $z \sim 2.15$).  Considering
galaxies above these mass limits we find that the masses of ``group''
galaxies are still  higher than those of ``field'' galaxies, for the lower bin in redshift ($P_{KS}=9.7 \cdot 10^{-4}$). At  $z > 1.2$, however, it is not possible to give a conclusive result
due to the low statistic caused by this mass cut.

Analogous results are found from the analysis of the luminosity
distribution of ``field'' and ``group'' galaxies. In particular, we
find that the distribution of galaxies in ``groups'' have on average brighter $M_I$
rest-frame magnitudes at all
redshifts. The results are similar also for the other rest frame
bands, implying that galaxies in high density environments have, on
average, greater bolometric luminosity with respect to field
galaxies.

\begin{figure}[!htb]
\resizebox{\hsize}{!}{\includegraphics{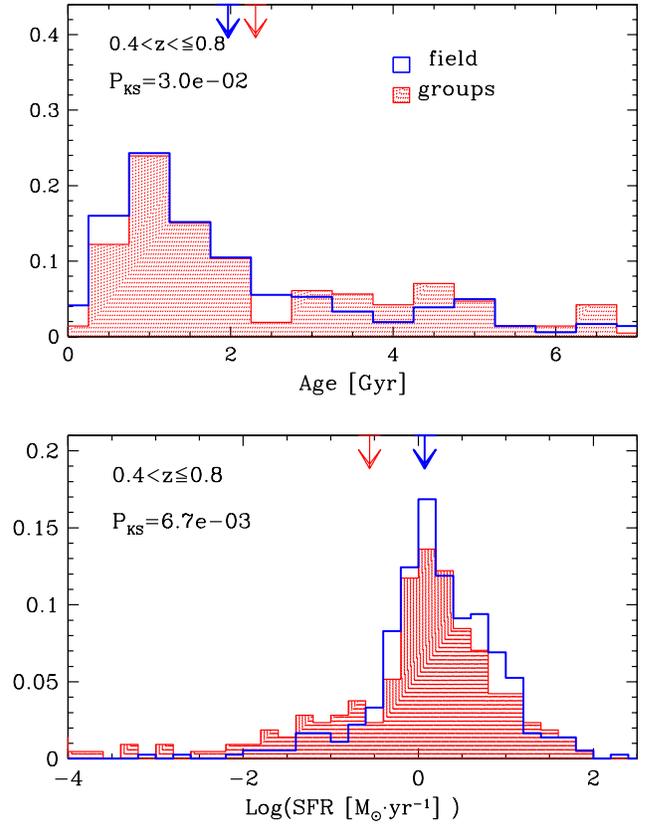}} \caption{top: As in Fig.
\ref{fig:mass}  but for the ages of galaxies. The average value of the age for the two
distributions are indicated by arrows. Bottom: As in Fig.
\ref{fig:mass} but for SFR of galaxies. The average values of the $\log(SFR)$ for the two
distributions are indicated by arrows.} \label{fig:age}
\end{figure}

Finally, we study the age  and SFR distributions for ``group'' and ``field''
galaxies (see Fig. \ref{fig:age}). Only at low redshift there appears to be a significant
difference (respectively $P_{KS}=3.0 \cdot10^{-2}$ and $P_{KS}=6.7 \cdot10^{-3}$, see Fig. \ref{fig:age}). The
two age distributions show a similar shape for young galaxies, but
``group'' galaxies have a higher fraction of old galaxies. As also
shown by the difference in the average ages for the two samples, the
``group'' galaxies are older than the field ones. At higher redshifts
the two distributions do not show significative differences. Indeed,
at higher redshifts, any possible difference in the age of the
two galaxy populations is probably smaller than the uncertainty on the
ages. Analogously, star forming galaxies have a similar distribution for ``group'' and ``field'' samples, but  the ``group'' sample has a higher  faction of galaxies with low star formation  as it is also shown by the different values of the average SFR.

\section{Summary and Conclusions}
We applied a (2+1)D algorithm on the GOODS-MUSIC catalogue to
identify structures in this area. This algorithm combines galaxy
angular positions and precise photometric redshifts to give an
adaptive estimate of the 3D density field effectively also at $z>1$,
and in a wide area. In this way we obtained a density map from
redshift 0.4 up to 2.5, and we isolated the higher density regions.
To identify density peaks we chose a conservative selection criteria (at least five galaxies in connected regions of $\rho >
\bar{\rho} + 4\sigma$) in order to maximise the purity of our
sample.

We built  mock catalogues simulating the GOODS-South
field. Applying our density thresholds and selection
criteria on these catalogues, we found a purity near  100\%  (with less than 15-20\% of
lost structures) up to redshift 1.8, and $\sim$ 75-80\% at higher
redshift.  In the higher redshift range, the criterium is very
conservative, to keep a low number of false detections, therefore
the completeness is low ($<40\%$). From the simulations we also
evaluated the ability of the algorithm in separating real structures
that are very close both in redshift and angular position. Both at
low and high redshift it is not possible to separate structures
whose centres are closer than $1.0 Mpc$ on the plane of the sky and
$2 \sigma_z$ in redshift. For larger separations it is possible to
distinguish the groups, but using higher thresholds (5 or 6 $\sigma$
above the average).

We found large scale overdensities at different redshifts ($\sim
0.6$, $\sim 1$, $\sim 1.61$ and $\sim 2.2$ ), which are well
traced by the AGN distribution, suggesting that the environment on
large scales ($\sim 10 Mpc$) has an influence on AGN evolution
\citep{silverman2007}. We
isolated several groups and small clusters embedded in these large scale structures. Most of the structures
at $z\sim 0.7$ and $\sim 1$ have properties of groups of galaxies: their
masses are of the order of $M_{200}= 0.2-0.8 \cdot 10^{14}
M_\odot$, and their X-ray luminosities are slightly below $10^{43} \
erg \ s^{-1}$, consistent with the expectations of the M$_{200}$-L$_X$ relations.  The structure at $z=0.71$, and those at $z>1.6$
seem to be more massive, and in particular the structures with  ID=4, and 9
can be classified as poor clusters. It is interesting to note that both these
 structures are significantly X-ray underluminous, as it is evident by a comparison with the M$_{200}$-L$_X$ relations found by \cite{reiprich2002} and by \cite{rykoff2008} (Fig. \ref{fig:x}). This is
not surprising since several authors have observed that optically
selected structures have an X-ray emission lower than what is
expected from the observations of X-ray selected groups and
clusters: this effect has been observed at low redshift both in
small groups \citep{rasmussen2006} and in Abell clusters
\citep{popesso2007} and in clusters at $0.6<z<1.1$
\citep{Lubin2004}. These results may be explained by the
fact that such optically selected structures are still in the
process of formation or the result of the alignment of two substructures along the line of sight, although it cannot be excluded that they
contain less intracluster gas than expected, because of the effect
of strong galactic feedback \citep{rasmussen2006}. If these
structures are virialised, as probable in the case of the massive structure at $z=0.71$
(ID=4), this may be an indication that they contain less
intracluster gas than expected. It is worth investigating this issue in future
deep surveys, since it would have interesting implications on the
evolution of the baryonic content of these structures.

We then studied the colour magnitude diagrams ($U-B$ vs $M_B$ ) for
all the structures. We defined the members of the red-sequence
according to the physical criterion $age/\tau \ge 4$ which should
select passively evolving galaxies with little residual star
formation. We confirmed no evolution of the red sequence slope up to
redshift $\sim 1$.
This implies that the mass-metallicity relation that produces
the slope of the red sequence remains constant up to $z \sim 1$.

We then studied the variation of the fraction of red and blue galaxies as
a function of the environmental density. We found that, at fixed
redshift, the red fraction increases at increasing $B$ luminosity,
while, at fixed luminosity, it increases with decreasing redshift.
We found that the increment of the red fraction at growing density
disappears at $z>1.2$. 

We also studied galaxy properties in different environments. We
found that the galaxies in high density environments have higher
masses with respect to ``field galaxies'', in qualitatively agreement
with a downsizing scenario. The mass distributions show
a significant difference in all but the last redshift bin.
Similarly, the galaxies in groups have on average brighter
rest--frame magnitudes and there is a greater number of bright
galaxies in groups at all redshifts compared to field galaxies.
Finally, the age and SFR distributions for the two subsamples appear
different only at low redshifts where ``group galaxies'' are generally
older and less star forming than``field'' ones.

From the analysis of the environmental
dependence of galaxy colours and mass as a function of redshift, and from the absence of any
well defined red sequence at high redshift, we can argue that
a critical period in which some basic characteristics of galaxy
populations are established is that between $z\sim 1.5$ and $z\sim
2$.


\end{document}